\documentclass{osa-article}

\journal{osajournal}
\articletype{Research Article}
\usepackage{siunitx} % CUSTOM DECLARATION added
\usepackage{amsmath,amssymb}
\usepackage{amsmath}
\usepackage{subfig}
\usepackage{graphicx}
\usepackage[utf8]{inputenc} % CUSTOM DECLARATION added

\newcommand{\pll}{{\mkern3mu\vphantom{\perp}\vrule depth 0pt\mkern2mu\vrule depth 0pt\mkern3mu}}
\newcommand{\vect}[1]{\ensuremath{\mathbf{#1}}} % CUSTOM DECLARATION applies vector representation in math mode

\begin{document}

%\title{Quantitative phase imaging on a commercial confocal microscope with sinusoidal synthetic optical holography}
%\title{Implementation of quantitative phase imaging in commercial confocal microscopes based on sinusoidal-wave synthetic optical holography}
%\title{Quantitative phase imaging on commercial confocal microscopes with sinusoidal-phase synthetic holography}
\title{Accessible quantitative phase imaging in confocal microscopy with sinusoidal-phase synthetic optical holography}
%\title{Synthetic Optical Holography (SOH) as an accessible Quantitative Confocal Phase Imaging technique}
%\title{An accessible implementation of Synthetic Optical Holography (SOH) in confocal systems}
%\title{Providing holographic capabilities to confocal systems: An accessible implementation}
%\title{An accessible implementation of Synthetic Optical Holography (SOH): Providing holographic capabilities to confocal systems}
%\title{Extending confocal microscopy with quantitative phase imaging based on sinusoidal synthetic holography}

\author{Arturo Canales-Benavides,\authormark{1} Yue Zhuo,\authormark{2,3,4} Andrea M. Amitrano,\authormark{5,6} Minsoo Kim,\authormark{5} Raul I. Hernandez-Aranda,\authormark{7} P. Scott Carney,\authormark{1,8} and Martin Schnell\authormark{2,*}}

\address{
\authormark{1}The Institute of Optics, University of Rochester. 275 Hutchison Rd, Rochester, NY 14620, USA\\
\authormark{2}Beckman Institute for Advanced Science and Technology, University of Illinois at Urbana-Champaign, Urbana, Illinois 61801, USA\\
\authormark{3}Department of Bioengineering, University of Illinois at Urbana-Champaign, Urbana, Illinois 61801, USA\\
\authormark{4}Micro and Nanotechnology Laboratory, University of Illinois at Urbana-Champaign, Urbana, Illinois 61801, USA\\
\authormark{5}Department of Microbiology and Immunology, David H. Smith Center for Vaccine Biology and Immunology, University of Rochester, USA\\
\authormark{6}Department of Pathology, University of Rochester Medical Center, University of Rochester, USA\\
\authormark{7}Tecnológico de Monterrey, Eugenio Garza Sada 2501 Sur, Monterrey, NL 64849, Mexico\\
\authormark{8}scott.carney@rochester.edu}
\email{\authormark{*}schnelloptics@gmail.com} %% email address is required



\begin{abstract}
We present a technically simple implementation of quantitative phase imaging in confocal microscopy based on synthetic optical holography with sinusoidal-phase reference waves. Using a Mirau interference objective and low-amplitude vertical sample vibration with a piezo-controlled stage, we record synthetic holograms on commercial confocal microscope (Nikon, model: A1R; Zeiss: model: LSM-880), from which quantitative phase images are reconstructed. We demonstrate our technique by stain-free imaging of cervical (HeLa), ovarian (ES-2) cancer cells and stem cell (mHAT9a) samples. Our technique has the potential to extend fluorescence imaging applications in confocal microscopy by providing label-free cell finding, monitoring cell morphology as well as non-perturbing long-time observation of live cells based on quantitative phase contrast.
\end{abstract}
\section{Introduction}

Digital holography\cite{RN7,RN66,RN217} is an enabling technique in light microscopy that provides new capabilities ranging from quantitative phase imaging (QPI) of unstained biological samples\cite{RN9,RN10} and acquisition of three-dimensional information\cite{RN11,RN12,RN13} to image correction based on numerical refocusing and aberration correction\cite{RN14,RN15}. The underlying basis is the superposition of light scattered from an illuminated object with a reference wave at the detector\cite{RN1, RN3, RN2}. The resulting interference pattern encodes the complex optical field of the object in form of a fringe pattern -- the hologram. Digital hologram recording with cameras enabled the reconstruction of amplitude and phase information in real-time and, together with its speed and ease of use, established holography as a tool in science and industry. 

Holography is commonly implemented in wide-field modalities, however, confocal microscopy offers a series of advantages over wide-field microscopy such as superior contrast, rejection of out-of-focus light and spectroscopic imaging capabilities\cite{RN219,RN220,RN221}. These advantages rendered confocal microscopy as an indispensable tool in biomedical imaging applications by enabling clear, high-contrast imaging of biological specimen, the acquisition of three-dimensional data in form of z-stacks and the simultaneous detection of multiple labels.  However, the integration of holography in confocal microscopy for quantitative phase imaging is widely unexplored. Holographic techniques using cameras instead of photodetectors -- the native device for light detection in confocal microscopy -- have been used. For example, quantitative phase imaging in a transmission-mode confocal microscope was demonstrated by interfering light collected from the sample with an off-axis reference wave on a line camera to implement parallel interferogram recording\cite{RN225}. Further, holographic detection of the object scattered light with a camera allowed for implementation of a virtual pinhole\cite{RN37,RN234} and for quantitative phase imaging\cite{RN235,RN243}. However, in all of these experiments, the phase was determined independently at each position of the sample rather than encoding the phase information across the image as in wide-field holography. Confocal phase measurement with photodetectors was demonstrated using interferometric detection schemes such as heterodyne\cite{RN226,RN223,RN202}, phase-locked\cite{RN227}, dual-phase\cite{RN228} and scanning interferometry\cite{RN146,RN206}, and optical metrology and vibration imaging have been demonstrated\cite{RN224}. Although offering straight forward phase imaging, the implementation of these modalities in existing confocal microscopes can be expected to be complex as dedicated components such as Wollaston prisms, acousto-optical modulators, modulated lasers or rapidly moving mirrors are required.
%Moreover, cameras perform generally inferior to photodetectors in terms of signal-to-noise ratio [refXXX PeterSo], spectral and frequency response. 
% optical surface profiling of micromechanical systems 
%\cite{Herring1997, Herring2004, Herring2005b}

Recently, synthetic optical holography (SOH)\cite{RN176} was introduced for quantitative phase imaging in near-field\cite{RN176}, confocal\cite{RN175} and scanning microcavity\cite{RN231} microscopy that take advantage of the mutual information between pixels and encodes amplitude and phase information across the image in the spirit of wide-field holography. In SOH, the scattered field from the focus or scanning probe is superposed with a reference wave with a linear-in-time phase function. Recording the detector signal pixel-by-pixel while scanning the focus or probe in raster fashion records a synthetic hologram from which amplitude and phase images can be reconstructed. SOH was first demonstrated for rapid phase image in near-field microscopy where the reference mirror vibration of pseudoheterodyne interferometry posed a speed limitation for the imaging process. By implementing SOH, the reference mirror vibration was replaced by a mirror movement at constant velocity, and imaging speed was improved by more than a magnitude\cite{RN176}. Further, SOH was shown to enable quantitative phase imaging in confocal microscopy for optical surface profiling with sub-nanometer vertical sensitivity\cite{RN175}, confocal hologram recording \cite{RN232}, fast quantitative phase imaging in a line-scanning modality\cite{RN229} and transient vibration imaging\cite{RN244}. These holographic as well as related interferometric techniques for phase retrieval were typically implemented on custom systems that afforded the necessary space and flexibility in reconfiguring the optical setup. However, modifications of commercial microscopes are often not allowed as a matter of policy, and hence quantitative phase imaging remained unavailable.  Given the large number of installed commercial confocal microscopes in research, imaging centers and industry, solutions that do not requiring any modifications to the microscope hardware could provide quantitative phase imaging capabilities to a large user base.

Here, we present quantitative confocal phase imaging based on SOH using sinusoidal-phase reference waves. We implement our technique in commercial confocal microscope systems (Nikon, model: A1R; Zeiss, model: LSM-880) by using only a Mirau interference microscope objective and low-amplitude vertical sample vibration with a piezo-actuated stage scanner. We show that this implementation is compatible with galvanometer-based beam scanning of the microscope, thus achieving fast imaging acquisition times at the full speed supported by the microscope. We demonstrate our technique by quantitative phase imaging of a test target and unstained cervical cells (HeLa), ovarian cancer cells (ES-2) and stem cells (mHAT9a).

\section{Implementation of sinusoidal SOH in a commercial confocal microscope}

\begin{figure}[!tb]
\centering
%\subfloat{\includegraphics[width=7.2cm]{Setup/SetupConfocal_v2}\label{galvo}}
%\\[-3ex]
%\subfloat{\includegraphics[width=4cm]{Setup/SetupModule1_v3}\label{os}}
%\subfloat{\includegraphics[width=3.5cm]{Setup/SetupModule2_v2}\label{zs}}
%\includegraphics[width=0.65\linewidth, trim=0in 1.9in 2.2in 1in, clip]{Setup/SetupConfocal_v6}
\includegraphics[width=0.65\linewidth, trim=0in 0.5in 0in 0.8in, clip]{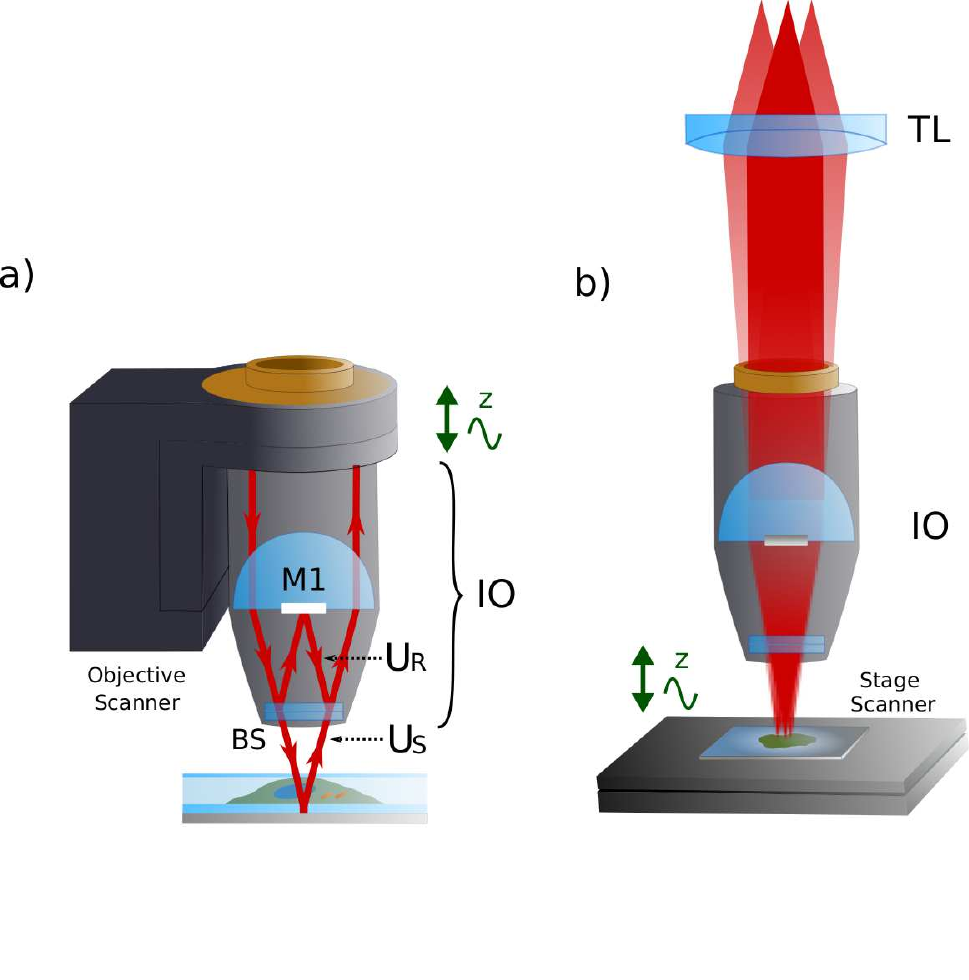}

\caption{Implementations of SOH phase imaging in a commercial confocal microscope. (a) Mirau interference objective mounted on a piezo-actuated objective scanner. Interference of the object scattered field $U_\text{S}$ with the reference field $U_\text{R}$ is depicted. (b) Fixed Mirau interference objective with sample placed on piezo-actuated stage scanner. Beam scanning operation is illustrated. Symbols: IO: Interference Objective, BS: internal beam splitter, M1: internal reference mirror. }
\label{Setup}
%The nanopositioning system is driven in a sinusoidal fashion while the sample is scanned by the galvo scanner. The interference intensity is detected pixel-by-pixel with the photodiode (PD) while out-of-focus light is rejected by the pinhole (P). By taking advantage of the mutual intensity information between pixels, a synthetic hologram is generated.
\end{figure}

Figure \ref{Setup} shows two possible implementations of SOH quantitative phase imaging in a commercial laser scanning confocal microscope. The basis is a Mirau interference objective to implement interferometric detection of the sample scattered light in compact form. The excitation light beam coming from the confocal microscope is split at a ratio of 50:50 at the internal beam splitter (BS) of the objective. The transmitted beam is focused on the sample, while the reflected beam is focused and reflected at the internal reference mirror (M1). At each position $\vect{r}=(x,y)$ on the sample, the scattered light from the sample  $U_\text{S}(\vect{r})=A_\text{S}(\vect{r})$ is recombined with the reference beam $U_\text{R}(\vect{r})=A_\text{R}$ at the beam splitter BS, where $A_\text{S}(\vect{r})$ is the complex scattered field from the sample and $A_\text{R}$ the spatially constant reference field. 

To recover quantitative phase information, we apply SOH with sinusoidal-phase reference waves, which are an example of a broader class of reference waves not normally found in digital holography, but which can be easily synthesized in SOH and offer technical advantages. This modality was originally demonstrated on a near-field microscope with an external reference arm and at infrared frequencies, where large images could be acquired with open-loop, limited travel range piezoactuators\cite{RN178}. The need for only small displacements of the reference mirror could allow for compact implementations of SOH in confocal microscopy. Indeed, previous implementations of SOH with linear-phase reference waves needed an external reference arm and a bulky, long-travel piezostage, and moreover only slow phase imaging owing to sample scanning was shown, while stain-less imaging of biological samples had not been demonstrated yet\cite{RN175}. In the following, sinusoidal-phase SOH is applied to confocal microscopy where the reference arm is compactly integrated in the microscope objective and galvanometer-based beam scanning provides for fast phase imaging. 

To generate sinusoidal-phase reference waves in a confocal microscope, either the Mirau interference objective (Fig. \ref{Setup}(a)) or the sample (Fig. \ref{Setup}(b)) is vertically vibrated with sinusoidal waveform. In both approaches, vertical vibration introduces a phase modulation in the sample beam of the interferometer. In a first approximation, we describe this phase modulation as linear with vertical sample displacement, taking the limit in vibration amplitude and objective NA. Thus, $\varphi_\text{mod}(t)=-\gamma \text{sin}(\omega t+\varphi_0)$, where $\gamma=2\pi s_0/\lambda$ is the modulation depth in radian and is determined by the vibration amplitude $s_0$ of the objective or stage scanner and wavelength $\lambda$ of the laser, $\omega$ the vibration frequency and $\varphi_0$ a phase offset generated when the vibration position zero does not coincide with scan position zero. The combined beam of the interferometer is detected by the detector unit of the confocal microscope. Hologram acquisition is performed by raster-scanning the focal spot using the microscope's xy-galvanometer scanner, producing the synthetic hologram, 
	\begin{equation}
	\label{eq:Udet}
		I(\vect{r}) = |U_\text{R} + U_\text{S} (\vect{r})|^2 = |A_\text{R} +  A_\text{S} (\vect{r}) e^{-i\gamma  \text{sin}(\vect{k}_\pll \cdot \vect{r}+\varphi_0)} |^2 ,
	\end{equation}
where the term $e^{-i\gamma  \text{sin}(\vect{k}_\pll \cdot \vect{r})}$ describes the reference field with a sinusoidal spatial variation of the phase with virtual wave vector $\vect{k}_\pll = (k_x, k_y)$ as a result of the sinusoidal phase modulation described by $\varphi_\text{mod}$\cite{RN178} . Taking a Fourier transform (FT), we have: 
	\begin{equation}
	\label{eq:UdetDFT}
		\tilde{I}(\vect{q}) = A_\text{R}^2\delta(\vect{q}) + |\tilde{A}_\text{S}(\vect{q})|^2 +  A_\text{R} \sum_{n=-\infty}^{\infty}    J_n(\gamma) e^{in\varphi_0}   \left[ \tilde{A}_\text{S}^*( n\vect{k}_\pll-\vect{q})  + (-1)^n \tilde{A}_\text{S}(  \vect{q} - n\vect{k}_\pll )  \right] ,		
	\end{equation}
where tilde indicates FT with respect to position and $J_n$ are Bessel functions. For even and odd values of $n$, the term in the square brackets can be seen to be proportional to the FT of the real and imaginary part of $A_\text{S}$, respectively\cite{RN178}. Reconstruction of the object scattered field is done in the following three steps. First, terms $n=1$ and  $n=2$ are isolated by applying a two-dimensional cosine window centered at position $-n\vect{k}_\pll$ in the FT space, shifted to the image center by $n\vect{k}_\pll$ and the inverse FT is calculated. Note that this step is identical to the reconstruction of synthetic holograms with linear-phase reference waves  $n\vect{k}_\pll$. Second, the reconstructed imaginary ($n=1$) and real parts ($n=2$) are combined to form the complex scattered field of the sample $A_\text{S}(\vect{r})$. Third, possible phase gradients owing to a sample tilt were removed by linear fit in $\vect{r}$, allowing for reliable phase imaging even in the presence of a slight tilt of the sample. Note that the wavevector $\vect{k}_\pll$ as well as the term weighting $J_2/J_1$ may be estimated by inspection of the FT of the hologram, $\tilde{I}(\vect{q})$, and thus robust reconstruction of amplitude and phase is possible even in cases where the piezo calibration is slightly off or where the piezo vibration amplitude and frequency is not precisely known. A Matlab code for reconstruction of sinusodial SOH holograms is provided in Code 1 (Ref. \cite{RN265}).

\section{Experimental demonstration}

\subsection{Sample preparation}
We employed focused-ion beam milling of a 40 nm thick Au film deposited on a $\text{CaF}_2$ substrate to fabricate a reflective 1951 USAF resolution test target of groups 8 and 9 (note: group 9 was mistakenly labeled '1'). The sample yielded highly reflective bars on a weakly reflective substrate (as shown in Fig. \ref{usaf}(c)), where the height difference between the bars and the substrate provides a phase contrast based on different optical path lengths of the probing beam. The ovarian cancer cells (Cellosaurus cell line, ES-2) and stem cells (murine dental epithelial stem cells, mHAT9a) were maintained in Dulbecco’s Modified Eagle Medium (DMEM) supplemented with 10\% Fetal bovine serum (FBS) and 1\% Penicillin Streptomycin  (PenStrep). The cells were cultured at a temperature of 37ºC and supplied with 5\% $\text{CO}_2$ humidified air environment. To allow for phase imaging in a reflection geometry, the cell samples were prepared on aluminum-coated (reflective) glass slides with a proprietary CHC thin film protective coating (Deposition Research Lab, Inc.; St. Charles, MO, USA). In more detail, the aluminum-coated glass slides were cleaned by sonication in isopropyl alcohol (IPA), acetone, and deionized (DI) water for one minute each, followed drying with nitrogen (N2) gas. Then, they were oxygen-plasma treated for further cleaning and to facilitate attachment of a liquid containment gasket formed from polydimethylsiloxane (PDMS). The aluminum-coated glass slide surface was hydrated with a phosphate buffered saline solution and coated with a thin layer of arbitrary ECM (e.g. fibronectin) to promote cellular attachment. The ovarian cancer cells and stem cells were transferred from the cell culture flask to the aluminum-coated glass slides (on the coating side, not on the opposite side of the slide) and cultured in the incubator for overnight. The cells were fixed with 4\% paraformaldehyde (PFA) and rinsed with 1x Phosphate-Buffered Saline (PBS) solution. A coverslip was then placed on the aluminum-coated glass slide with residual PBS serving as medium between the coverslip and the glass slide and subsequently secured using clear nail polish. The HeLa cells were plated on a Poly-L-lysine coated glass coverslip, incubated overnight and fixed with a 3.7\% paraformaldehyde solution in DMEM media the following day. Following fixation, the cells were washed in PBS, the coverslip was placed on an aluminum-coated glass slide (on the side of the coating) with residual PBS serving as medium and secured using clear nail polish.
% The coverslip was then placed on an aluminum-coated glass slide (on the side of the coating) with residual PBS serving as medium between the coverslip and the glass slide and subsequently secured using clear nail polish.

\subsection{Instrument setup}
We implemented our method in a commercial confocal microscope (Nikon, model: A1R) following the approach of sample vibration (Fig. \ref{Setup}(b)). We used a 20x Mirau interference microscope objective (Nikon, model: CF IC EPI Plan DI 20x) with 0.4 NA and a parfocal length extender (Thorlabs Inc., model: PL15RMS) to match the parfocal length of the other installed objectives. Vertical sample vibration was realized with a piezo-actuated nanopositioner (Mad City Labs Inc., model: Nano-Z100) that was interfaced with a waveform generator (Keysight, model: 33512B). The nanopositioner was a drop-in replacement for the existing slide holder. No modifications to the microscope hardware were required during installation. The optical path was set up for imaging of the elastically scattering light in reflection. In detail, laser illumination at $\lambda = \SI{561}{\nano\meter}$ wavelength was chosen for SOH imaging. A non-dichroic beam splitter was selected to allow for both transmission and reflection of the SOH imaging wavelength (labeled ``BS 20/80`` in the optical path settings). Further, open pinhole settings were typically used ($\SI{255}{\micro\meter}$), all filters were removed in the detector unit (e.g. by selecting ``Through``  in the optical path dialogue) and a channel with normal photomultiplier tube was selected for photo detection (e.g. channel 4). To optimize spatial resolution and signal-to-noise ratio in the reconstructed phase images, the vibration amplitude was set to $s_0 = \gamma \lambda / 2\pi = \SI{0.23}{\micro\meter}$ such that $\gamma=2.63$ and hence $J_1 = J_2$ \cite{RN45}, yielding equal weighting of the imaginary and real terms of the object scattered field $A_\text{S}(\vect{r})$. Further, the vibration frequency $\omega$ was  set such that terms 1 - 3 were distributed across the entire FT space. The adjustment of both vibration amplitude and frequency was verified by imaging an empty space on the sample (which effectively acted as plain mirror) and subsequent determination of term position and weighting in the Fourier transform of the hologram. Cell imaging was performed by focusing on the aluminum substrate. As light traversed the sample, it accumulated a phase shift owing to the higher refractive index of the cellular components in comparison to the mounting medium (illustrated in Fig. \ref{Setup}(a)), which provided a non-specific, label-free map of (sub)cellular structure. Note that our method is alignment-free (no adjustments needed to be made to the Mirau interference objective) and calibration-free (a single-shot hologram acquisition of the sample was sufficient as the optical phase was directly measured, prior acquisition of calibration data was not needed), which reduces the complexity of phase imaging for the user.
%at a power level of about $2\%$

%Prior to imaging and with the sample removed, the collar of the 50x Mirau interference objective was adjusted for maximum signal (given by the reflection from the internal reference mirror), while no adjustment was needed in case of the 20x objective. 

\subsection{Results}
\begin{figure}[!tb]
    \centering
  %  \subfloat{\includegraphics[width=4.3cm]{Images/cancerphase}\label{cancerphase}}
  %  \hspace{0.1cm}
  %  \subfloat{\includegraphics[width=4.3cm]{Images/cancerphasezoom}\label{cancerphasezoom}}
    %\\
   % \subfloat{\includegraphics[width=4.3cm]{Images/stemphase}\label{stemphase}}
   % \hspace{0.1cm}
   % \subfloat{\includegraphics[width=4.3cm]{Images/stemphasezoom}\label{stemphasezoom}}
    \includegraphics[width=0.6\linewidth, trim=0in 0in 0in 0in, clip]{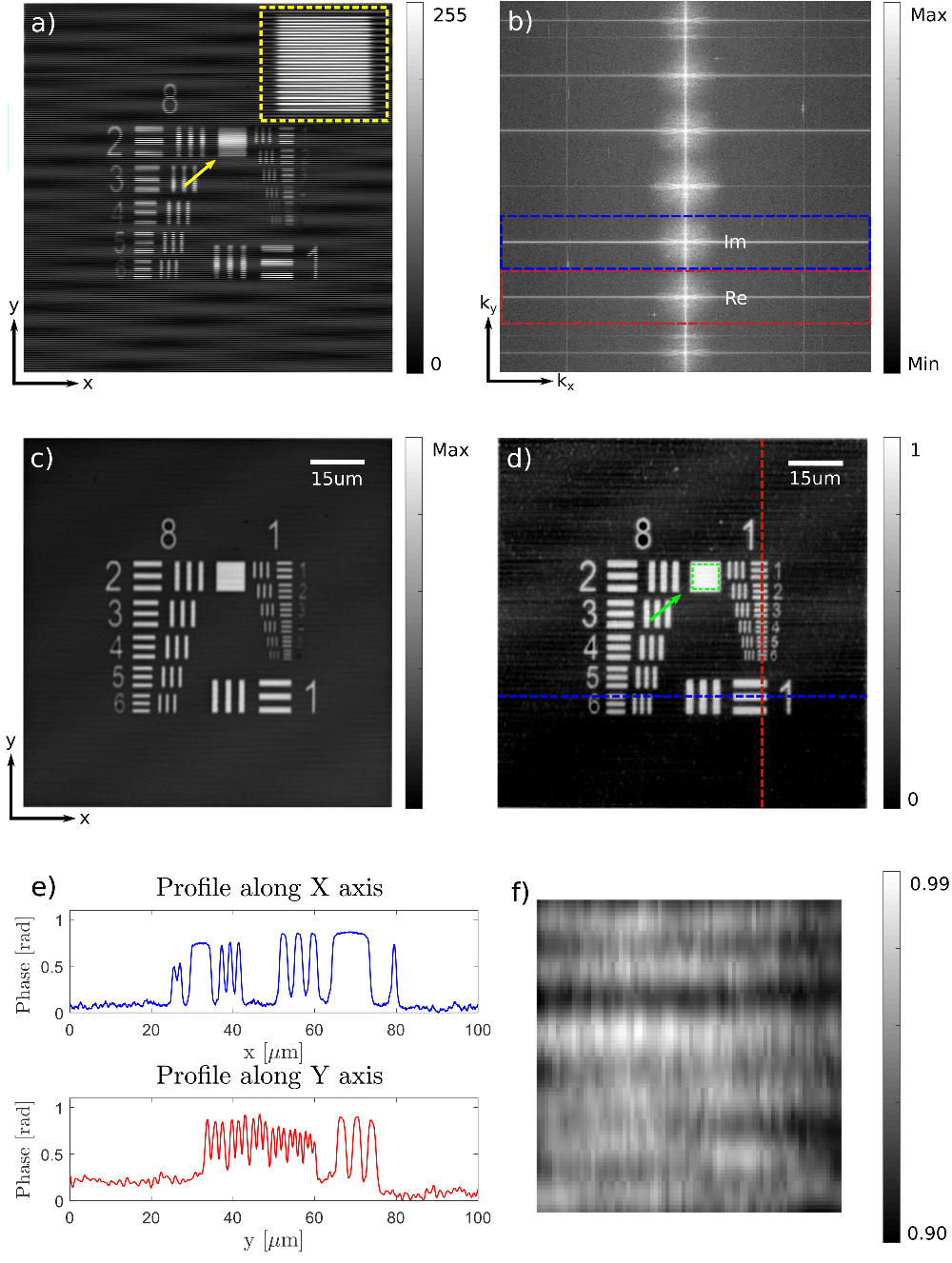}
    \caption{Quantitative confocal phase imaging of a test target. (a) Synthetic Hologram, $I(\vect{r})$,  zoom on square of group 8 (inset). Hologram data is displayed in analog-digital unit (ADU). (b) Two-dimensional Fourier Transform of (a), $\tilde{I}(\vect{q})$. Dashed boxes indicate window size in the reconstruction of the imaginary and real part of the scattered light from the sample, $A_\text{S}$. (c,d) Reconstructed amplitude and phase images of $A_\text{S}$. Phase is displayed in radians. (e) Line profiles of the reconstructed phase taken along the horizontal and vertical dashed lines in (d). (f) Zoom on square of group 8 in (d), illustrating spatial phase noise. }
    \label{usaf}
\end{figure}

We first demonstrated our method by quantitative phase imaging of the test target in Fig. \ref{usaf}. To record the synthetic hologram, we supplied a sine waveform at frequency $\omega=40 \text{Hz}$ and with 177 mV amplitude into the control port of the supplied sample stage controller, yielding completion of one sample vibration period in about 6 scan lines of the hologram with an estimated peak-to-peak amplitude near $2s_0 = \SI{0.46}{\micro\meter}$ to achieve the required modulation depth of $\gamma=2.63$. The acquired synthetic hologram exhibits a dense, regular fringe pattern (Fig. \ref{usaf}(a)) with strong and weak fringe contrast on the highly reflective Au bars and on the weakly reflective substrate, respectively. FT of the hologram reveals nearly vertical arrangement of the Fourier terms 1 - 3 that are equally spaced at multiples of $\vect{k}_\pll$ from the center of the FT (Fig. \ref{usaf}(b)). The dashed boxes indicate the size of the two-dimensional cosine window used for isolation of terms $n=1$ and  $n=2$. The reconstructed amplitude and phase images in Figs. \ref{usaf}(c) and (d) show the local reflection amplitude and phase contrast introduced by reflection of the focus at the higher surface of the bars in comparison to substrate as well as the material-dependent phase shift on reflection. All elements of group 8 and 9 were resolved in the phase image and line profiles (Fig. \ref{usaf}(e)), indicating a spatial resolution of about $\SI{1.1}{\micro\meter}$. Phase sensitivity was estimated by calculating the spatial phase noise on a flat section of the sample, the square of group 8 (Fig. \ref{usaf}(f)). We obtained 13 mrad RMS, which is equivalent to a detection sensitivity of $\lambda/483$ in the optical path length. The observed spatial variations of the optical phase can be attributed to fluctuations in the optical path length of the interferometer, e.g. externally-induced mechanical vibration of the Mirau objective with respect to the sample along the z-axis and air turbulence in the unshielded microscope setup. Indeed, the presence of faint fringes in the direction of the fast scanning axis (horizontal fringes) show that phase noise is dominated by low frequency components, which indicates that interferometer instabilities are likely the limiting factor in phase sensitivity. Given that the mechanical structure of commercial confocal systems is not optimized for phase imaging, this is still a remarkable result and clear phase images of biological samples can be provided.
%In more detail, we chose the square of group 8 to determine phase sensitivity as this part of the sample constitutes the non-milled part and thus does not suffer from additional milling-induced surface roughness as it is the case with the substrate (see Fig. \ref{usaf}(e)). The RMS value was calculated by integration across the individual pixels composing the square of group 8.

%A horizontal line profile quantifies the induced phase shift to be $\approx 0.8$ rad.
% RMS value of vertical sensititivy?

\begin{figure}[!tb]
    \centering
    \includegraphics[width=0.6\linewidth, trim=0in 0in 0in 0in, clip]{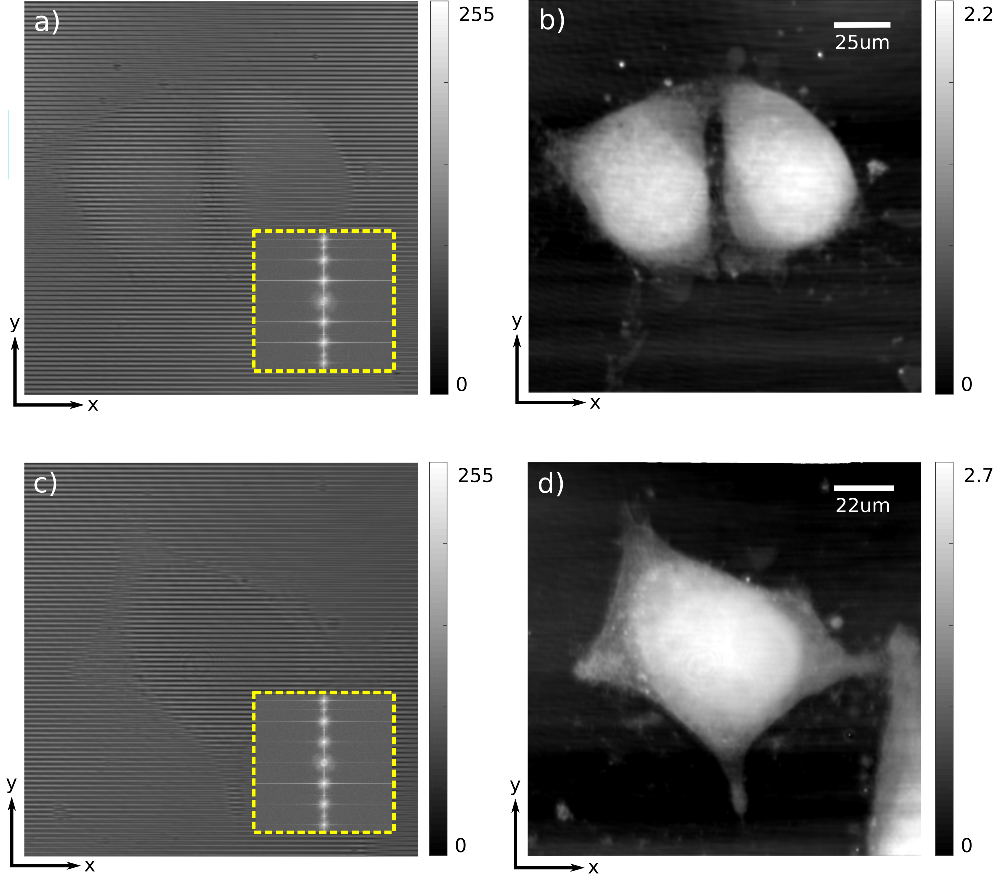}
    \caption{Quantitative confocal phase imaging of fixed cervical cancer cells (HeLa). (a,c) Synthetic Holograms and Fourier Transform (inset). (b,d) Reconstructed phase images (displayed in radians). }
    \label{HeLa}
\end{figure}

%Next, we applied our method to quantitative phase imaging of HeLa cells, ovarian cancer cells (ES-2) and stem cells (mHAT9a). In Figs. \ref{HeLa}(a) and (b) we show synthetic holograms and the corresponding FT of HeLa cells. The reconstructed phase images (Figs. \ref{HeLa}(c) and (d)) reveal cell shape and sub-cellular features based on an endogenous mass-based contrast provided by the optical phase delay imparted on the probing light beam as it traverses the cell. In more detail, the edges of the adhered cells are clearly reproduced by the phase images. The bright round central region (i.e. region of large optical phase delay) indicates the location of the cell nucleus and dense organelles surrounding the nucleus are also visible. Although not optimized yet, these images show that our technique enables basic morphology analysis by non-specific phase imaging that is sensitive to both labeled and unlabeled structures. 
Next, we applied our method to quantitative phase imaging of cervical cancer cells (HeLa), ovarian cancer cells (ES-2) and stem cells (mHAT9a) as examples for cells typically used in research. In Figs. \ref{HeLa}(a) and (c) we show synthetic holograms and the corresponding FTs of HeLa cells. The reconstructed phase images (Figs. \ref{HeLa}(b) and (d)) reveal cell shape and sub-cellular features based on an endogenous mass-based contrast provided by the optical phase delay imparted on the probing light beam as it traverses the cell. In more detail, the edges of the adhered cells are clearly reproduced by the phase images. The bright round central region (i.e. region of large optical phase delay) indicates the location of the cell nucleus and dense organelles surrounding the nucleus are also visible.

%(a) nd992.nds

\begin{figure}[!tb]
    \centering
    \includegraphics[width=0.9\linewidth, trim=0in 0in 0in 0in, clip]{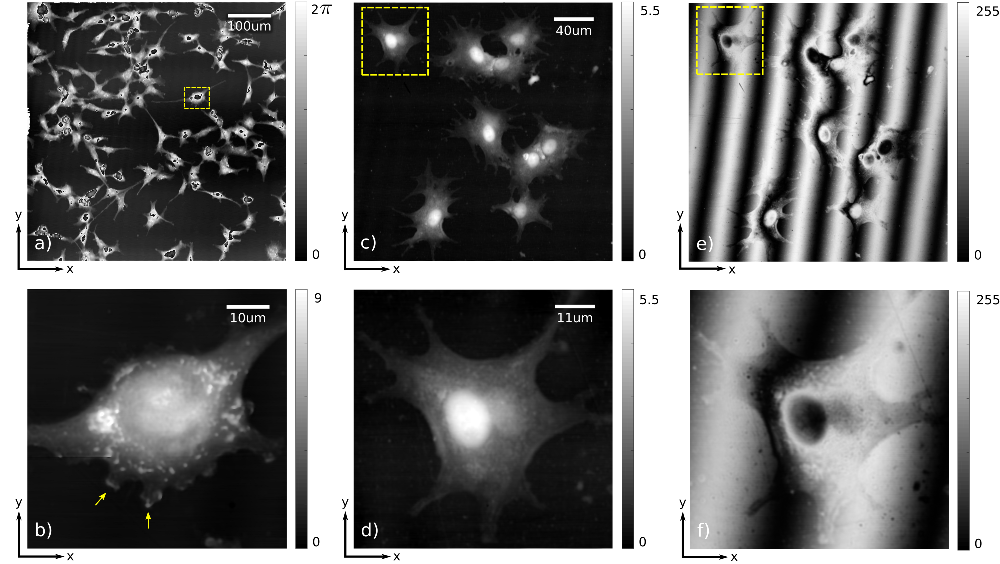}
    \caption{Quantitative confocal phase imaging of fixed ovarian cancer cells (ES-2) and Murine dental epithelial stem cells (mHAT9a). (a,b) Overview and high-resolution images of cancer cells.  (c,d) Overview and high-resolution images of stem cells.  (e,f) Homodyne phase-contrast images of the stem cells as obtained without SOH.}    
    \label{cancerstem}
\end{figure}

%To further demonstrate the utility of confocal phase imaging in biological imaging applications, in Fig. \ref{cancerstem} we applied our technique for rapid, non-perturbing large area imaging of cell samples. Figure \ref{cancerstem}(a) shows an overview image of an ovarian cancer cell sample (ES-2) that covered an area of $\SI{606 x 606}{\micro\meter}$ at 2048 x 2048 pixel and was acquired at the full speed of the confocal microscope in \SI{8}{\second} time (sample was vibrated at $\omega \approx 30 \text{Hz}$ at 40 mV). For the purpose of illustrating cell morphology inspectio , we re-imaged a single cell at higher resolution in Fig. \ref{cancerstem}(b) ($\SI{60 x 60}{\micro\meter}$, phase unwrapping applied). In this image, the cell nucleus can be recognized as a large oval area in the cell center exhibiting strong phase delay (bright color), as well as the presence of lamellipodia indicating migration capability of the cancer cell. Interestingly, bright spots are observed on the tips of most of the filopodia and lamellipodia (indicated by arrows), which could potentially be assigned to focal adhesion by comparison with fluorescence images of the same cell line\cite{RN239,RN238,RN242}, however, specific determination with e.g. fluorescent dyes is needed for unambiguous identification. 
To further demonstrate the utility of confocal phase imaging in biological imaging applications, in Fig. \ref{cancerstem} we applied our technique for rapid large area imaging of cell samples. Figures \ref{cancerstem}(a) and (c) show overview phase images of ovarian cancer (ES-2) and stem cells (mHAT9a), covering $\SI{606 x 606}{\micro\meter}$ and $\SI{260 x 260}{\micro\meter}$, respectively. Both images were acquired at 2048 x 2048 pixel at the full speed of the confocal microscope in \SI{8}{\second} time. For hologram generation, the samples were vibrated at $\omega \approx 30 \text{Hz}$ at 40 mV. The images reveal position and shape of several cells. Exemplarily, we re-imaged a single cell at higher resolution (Figs. \ref{cancerstem}(b) and (d)). The cell nucleus can be clearly recognized as a large oval area in the cell center exhibiting strong phase delay (bright color). The stem cell further shows a distribution of faint bright spots that we attribute to small organelles (Fig. \ref{cancerstem}(d)). The phase image of the cancer cell reveals the presence of lamellipodia (Fig. \ref{cancerstem}(b)). Interestingly, bright spots are observed at the tips of most of the lamellipodia (indicated by arrows), which might be assigned to focal adhesion by comparison with fluorescence images of the same cell line\cite{RN239,RN238,RN242} and would indicate migration capabilities of the cell. However, specific determination of the nature of these spots as well as the bright spots surrounding the cell nuclei with e.g. fluorescent dyes is needed for unambiguous identification. Although not optimized yet, these images show that our technique enables basic morphology analysis by non-specific phase imaging that is sensitive to both labeled and unlabeled structures. 

In comparison, phase-contrast-only images as obtained by confocal imaging with a Mirau interference objective, but without applying SOH, yielded unreliable image contrast. The latter was produced by a periodic signal modulation across the image that was introduced by a slight sample tilt (Figs. \ref{cancerstem}(e) and (f)). This modulation led to contrast inversion, and as a result cellular structure such as cell nuclei appear both with positive (white) and negative (black) signal, making cell finding and morphology analysis difficult. Such sample tilts could easily be corrected in the quantitative phase images by subtraction of a linear phase gradient, resulting in reproducible phase contrast (Figs. \ref{cancerstem}(a)-(d)). Importantly, such phase imaging at the longer wavelength of 561 nm is effective at locating and identifying individual cells based on their general morphology while reducing fluorophore excitation and possible photo bleaching and toxicity in comparison to a fluorescence-based imaging approach. 

% yue has seen a lot of fluorescence iamges and thus recognizes the fa
% we can take fluorescence image of ES-2, yue could also do SOH with confocal microscopy []

% fa have certain size, small typically less than 1um, orientation in migration direction, not random; 
% alternatively could be small organelles, but ned fluoroescnce to find out for sure
% fa apear bright because: fa = protein cluster formed at cell surface, yielding higher mass density (not exactly same, related to optical density, maybe n=1.4..1.5 ).

% stem should also have fa, but cancer cells show have strongest fa, so maybe less obvious with fa

% stem cells have sometimes two nuclei on bad split, but this would usually yield a stretched out cell
% stem cells are 2.5 time larger than cancer; stem ~30..40um diameter

% - why show FA up with an optical phase delay?
% - what is the reason that the cancer cell shows FA?
% - why only ES-2, but not the stem cells show FA so clearly?
% - explanation of shade of a second nulceus in stem cell image
%, surrounded by a number of bright spots that we attribute to small organelles distributed throughout the cytoplasm, which is typically expected in cancer cell cytoplasm (RefXXX). Further, bright spots are observed on the tips of most of the filopodia and lamellipodia, which can be assigned to focal adhesions and may indicate strong migration capability of the cancer cell (RefXXX).

%original data
%(a) 06162017_Cancer2_30Hz_40mV.nd2
%(b) 06162017_Cancer2_Zoom_30Hz_40mV.nd2
%(c) 06162017_Stem5_30Hz_40mV.nd2
%(d) 06162017_Stem5_Zoom_30Hz_40mV.nd2
%(f) 06162017_Stem5_Zoom_0Hz_0mV.nd2
%overview cancer: 606 x 606 um; zoom cancer: 60 um x 60 um
%overview stem: 260 x 260; zoom stem 70 um x 70um
%06162017_Cancer2_30Hz_40mV_3
%Phase_06162017_Cancer2_Zoom_30Hz_40mV_3.tif
%06162017_Stem5_Zoom_30Hz_40mV.nd2
%Phase_06162017_Stem5_Zoom_30Hz_40mV_3.tif
% all of them 8sec
%

\section{Discussion}
We tested our implementation of SOH on another commercial confocal microscope (Zeiss, model: LSM-880). We obtained phase images of stem cells of similar quality to those taken with the Nikon microscope (model: A1R), see appendix. Conveniently, both the Nikon and Zeiss systems were already equipped with a nanopositioning Z stage, thus only the Mirau interference objective and a waveform generator were required for SOH. In principal, any confocal system with a nanopositioning Z stage that can be user controlled during image acquisition is amenable to a similar modification. For better phase imaging performance, it is recommendable that (a) a non-dichroic beam splitter is installed in the microscope for efficient collection of the reflected light, (b) the microscope setup is floated on an optical table for vibration isolation and (c) acoustic noise in the laboratory is reduced to normal levels to avoid introduction of excessively large noise in the phase image. In the current implementation, data were recorded with the microscope software, saved to disk and reconstructed in post-process in Matlab. For live phase imaging capabilities, reconstruction of the phase image could potentially be implemented in the microscope software as only 2D Fourier transforms and simple matrix manipulations are required. Execution time of our reconstruction algorithm - implemented in Matlab, run on a modern workstation computer and not further optimized - was $\SI{0.8}{\second}$ for an image size of 2048 x 2048 pixels, potentially allowing for live phase reconstruction as the hologram is being acquired.

We note that the presented phase images appear free of halo artifacts. Halo artifacts are typically observed in wide-field phase contrast and some common-path QPI modalities and produce negative perimeter of an object akin to high-pass filtering\cite{RN240,RN241}. They were attributed to the limited spatial coherence of the illuminating field since the reference field is derived by spatial filtering of the sample field. In our data, the absence of halo effects can be recognized with the test sample shown in Fig. \ref{usaf}(d), where the line profile extracted from the phase image does not show any negative dimples around the bars. Similarly, visual inspection of the substrate close to cell boundaries does not reveal halo-typical local darkening in comparison to far-away locations on the substrate (Figs. \ref{HeLa} and \ref{cancerstem}). We explain these observations by the direct derivation of the reference field $U_\text{R}$ from the incident beam as well as by the confocal arrangement which provides spatial filtering of multiple scattered light that would normally contribute to the halo effect in wide-field modalities. Unambiguous observation of small objects and features at boundaries such as the structure of cell edges is thus possible with our method.

Sinusoidal SOH is a holographic method for phase retrieval that relies on spatial filtering and thus requires some degree of oversampling in the slow-scanning (y) direction, as determined either by the inherent bandwidth of the sample or the confocal microscope. In the high-resolution images of single cells (e.g. in Figs. \ref{cancerstem} (b) and (d)), we obtained good results when imaging an area of $\SI{173 x 173}{\micro\meter}$ (zoom setting 3.5 in the Nikon A1R) at an image size of 2048 x 2048 pixel, thus yielding an oversampling factor of $\sim 6$ for compensating the spatial filtering in the reconstruction process with window size of 1/6 of the bandwidth of the image. Such oversampling leads to an increase in imaging time in comparison to non-holographic imaging, however, since imaging could be done at the full speed of galvanometer scanner, we still obtained reasonable short imaging times of $\SI{8}{\second}$ with SOH. Realizing that all odd and all even terms carry the same information, advanced reconstruction methods could allow to make better use of the FT space and help to achieve lower oversampling factors. 
%This performance is two to three magnitudes faster than our previous work on confocal SOH phase imaging based on a sample scanning approach (ref. XXXX). 
%For taking overview phase images (Figs. \ref{cancerstem}(a) and (c)), lower oversampling factors and some term overlap may be tolerated because the image content of natural samples is typically concentrated at low spatial frequencies. 

For properly oversampled holograms, the spatial resolution is mainly determined by the numerical aperture of the microscope objective as in normal microscopy. At the low NA of our Mirau objective of only 0.4, the spatial resolution was estimated to be of the order of $\SI{1.1}{\micro\meter}$. Mirau interference objectives with higher NA are commercially available, promising phase imaging with correspondingly higher spatial resolution. Relatedly, interferometric confocal microscopy intrinsically provides depth sectioning capabilities\cite{RN222,RN262}. However, sectioning at the subcellular scale will require a much higher NA than demonstrated here and remains to be explored in future work. We note that regular imaging operation of the confocal microscope could easily be restored by turning off vibration of the piezo stage, switching to a different microscope objective and reconfiguration of the optical path (filters, pinhole), all of which can be done automatically on modern confocal systems without requiring manual intervention. This enables phase and fluorescence imaging on the same instrument and points to new imaging procedures where phase imaging is first used to locate cells and obtain non-specific information on cell morphology, after which high-resolution and depth resolved fluorescence imaging could be applied with high-NA objectives to reveal labeled structures of the cell with high specificity.

Further, we tested phase imaging at several wavelengths simultaneously as provided by the microscope (\SI{487}{\nano\meter}, \SI{513}{\nano\meter} and \SI{561}{\nano\meter}) and obtained good results (see appendix). In principle, the imaging wavelength can be freely selected within the specifications of the Mirau interference objective which allows the use of long wavelengths where fluorophore excitation and thus photo bleaching of the sample could be reduced.
%With the 20x Mirau interference objective, phase-imaging with sufficient signal-to-noise (SNR) ratio was possible at low laser power settings in the range of XXXX $2\%$, hence little thermal load on the sample when phase imaging. However,
%Fluorescence imaging was limited in terms of SNR and spatial resolution which we attribute to the low light collection efficiency of the interference objective as a result of (a) the low objective NA , (b) the internal beam splitter with only nominal 50\% transmission, (c) chromatic aberration yielding different focal planes between the fluorescence and SOH imaging wavelengths. 

\section{Conclusion}
We have presented quantitative phase imaging with two commercial confocal microscopes (Nikon, model: A1R; Zeiss, model: LSM-880) based on synthetic optical holography with sinusoidal-phase references waves. We demonstrated label-free, confocal phase imaging of cervical (HeLa), ovarian cancer and stem cells. We showed that large-area phase images could be useful for localizing cells while high-resolution phase images provided non-specific information on the cell morphology. Implementation of our method only required a Mirau interference objective and low-amplitude vertical sample vibration by means of a piezo-actuated stage scanner, while opening and modifications of the microscope were not needed. Particularly, the implementation did not interfere with regular microscope operation. This significantly facilitates implementation of quantitative phase imaging in existing, serviced commercial confocal setups. Image acquisition was done with the user-friendly software interface provided with the microscope. Real time reconstruction of the phase images could be implemented in the same software for realizing real-time phase imaging.

\section*{Appendix A: Implementation of SOH in a Zeiss confocal microscope}
\begin{figure}[h]
    \centering
    \includegraphics[width=0.9\linewidth, trim=0in 0in 0in 0in, clip]{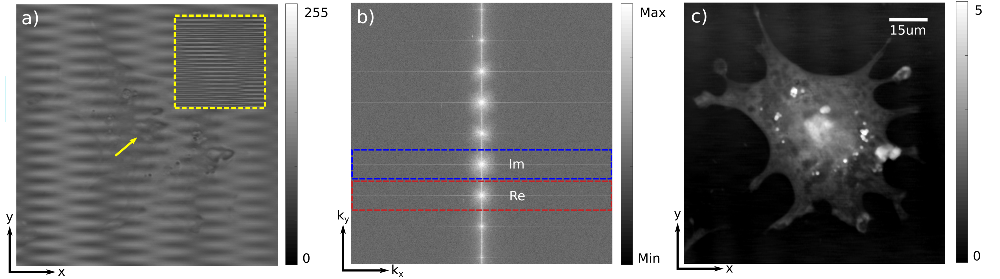}
    \caption{Demonstration of SOH on another commercial confocal system (Zeiss, model: LSM-880).  (a) Synthetic Hologram, $I(\vect{r})$, and zoom to reveal fringes (inset) (b) Two-dimensional Fourier Transform of (a), $\tilde{I}(\vect{q})$. Dashed boxes indicate window size in the reconstruction of the imaginary and real part of the scattered light from the sample, $A_\text{S}$. (c) Reconstructed phase image of a stem cell (unit: radians). } 
    \label{zeiss}
\end{figure}

We have implemented SOH in other commercial confocal systems and applied it to phase imaging of a stem cell. Fig. \ref{zeiss} shows data obtained with a microscope from Zeiss (model: LSM-880). FT of the hologram (Fig. \ref{zeiss}(b)) shows the expected distribution of the individual terms, enabling the reconstruction of a quantitative phase image  (Fig. \ref{zeiss}(c)), as described above.
% piezo stage make and model? microscope objective?

\section*{Appendix B: Simultaneous multicolor phase imaging}
\begin{figure}[h]
    \centering
    \includegraphics[width=0.9\linewidth, trim=0in 0in 0in 0in, clip]{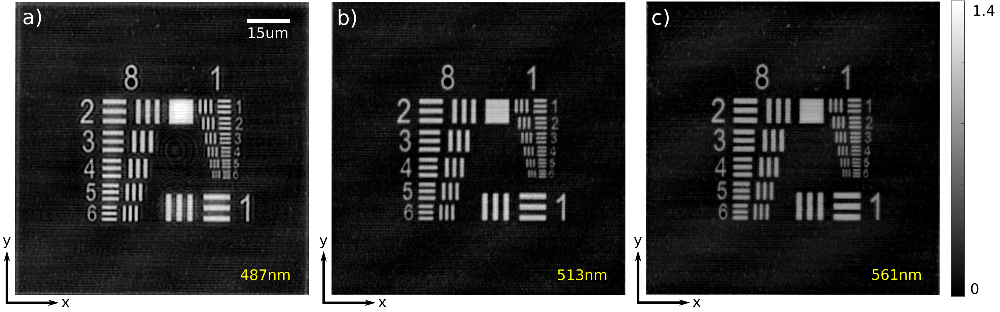}
    \caption{Simultaneous multicolor phase imaging with SOH on a commercial confocal system (Nikon, model: A1R). (a-c) Phase images obtained of the test target in Fig. \ref{usaf} at 487, 513 and 561 nm. Phase is displayed in radians.} 
    \label{multicolor}
\end{figure}

SOH allows for phase imaging at different wavelengths with a single reference arm and simultaneously. In the first implementation of this modality, wavelength separation was provided externally by a wavelength-dependent detector responsivity\cite{RN176}. In the following, we show that multicolor SOH is possible with a commercial confocal system (Nikon, model: A1R) by making use of the integrated spectral detector unit. We employed sample illumination at three laser wavelengths simultaneously, $\lambda = \{487, 513, 561\}\text{nm}$ and used channels 1 (PMT), 2 (PMT GaAsP) and 4 (PMT) for detection. For wavelength separation, the following filters were used: 485/30 (center wavelength/width) on channel 1, 525/50 on channel 2 and no filter on channel 4. Sample vibration was done at frequency $\omega=41 \text{Hz}$ and with 177 mV amplitude, similar to above experiments. The reconstructed phase images are shown in Fig. \ref{multicolor}.  Slightly larger phase contrast was observed at the shorter wavelength, attributable to the larger height-induced phase change and different material-specific reflection phase. Simultaneous multicolor phase imaging could be applied to resolve the $2\pi$ ambiguity of single-wavelength phase imaging \cite{RN260,RN261}.

\section*{Appendix C: Acquisition parameters}
The following acquisition parameters were used in Fig. \ref{usaf}: 2048 x 2048 pixel image size, \SI{8}{\second} acquisition time using unidirectional scanning, laser power $4.5\%$, pinhole size $\SI{255.4}{\micro\meter}$ (14.9 Airy units (AU)), PMT high voltage $38$ and PMT offset setting $0$.  Original image dimension is \SI{151 x 151}{\micro\meter}, panels (a,c,d) shows zoom into data at \SI{100 x 100}{\micro\meter}.

The acquisition parameters for Fig. \ref{HeLa} were: 2048 x 2048 pixel image size, \SI{32}{\second} acquisition time using unidirectional scanning, laser power $5\%$, pinhole size $\SI{58.7}{\micro\meter}$ (3.4 AU), PMT high voltage $45$ and PMT offset setting $25$. Image dimensions are \SI{151 x 151}{\micro\meter}. 

The acquisition parameters for Fig. \ref{cancerstem} were: 2048 x 2048 pixel image size, \SI{8}{\second} acquisition time using unidirectional scanning. Laser power, pinhole size, PMT high voltage and PMT offset: $(20\%, \SI{31.9}{\micro\meter} (1.8 \textrm{AU}), 25, 10)$ in panel (a), $(20\%, \SI{31.9}{\micro\meter} (1.8 \textrm{AU}), 24, 5)$ in (b), $(26\%, \SI{24.3}{\micro\meter} (1.4 \textrm{AU}), 24, 5)$ in (c,e),  $(26\%, \SI{24.3}{\micro\meter} (1.4 \textrm{AU}), 23, 5)$ in (d,f). Image dimensions are \SI{606 x 606}{\micro\meter} in (a), \SI{260 x 260}{\micro\meter} in (b), \SI{60 x 60}{\micro\meter} in (c,e), \SI{70 x 70}{\micro\meter} in (d,f).

The acquisition parameters for Fig. \ref{zeiss} were: 2048 x 2048 pixel image size, \SI{5}{\second} acquisition time using unidirectional scanning, imaging wavelength $\SI{488}{\nano\meter}$, pinhole was opened, sample was vibrated at a frequency of $\omega=48 \text{Hz}$ and with 55 mV amplitude (using 50 Ohm termination at the piezostage controller). Similar to the Nikon microscope, a 80:20 beam splitter was used in the beam path of the Zeiss (labeled MBS T80/R20).  Detector CH1 was used for light detection. To detect the elastically scattered light, the spectral filter needed to be set in the following way: the left edge was set exactly on the laser wavelength, the right edge was set to the longest wavelength. Typical values for laser power, detector gain, offset and digital gain were 15\%, 270, 0 and 1.

% laser power $4.5\%$, pinhole size $\SI{255.4}{\micro\meter}$, PMT high voltage $38$ and PMT offset setting $0$.  Original image dimension is \SI{151 x 151}{\micro\meter}, panels (a,c,d) shows zoom into data at \SI{100 x 100}{\micro\meter}.

%2048 x 2048 pixels. Acquisition time = 5 sec. Wavelength = 488 nm. Pinhole open; SOH parameters: 48 Hz, 55 mV. 

The acquisition parameters for Fig. \ref{multicolor} were: 2048 x 2048 image size, \SI{8}{\second} acquisition time using unidirectional scanning. Laser power, pinhole size, PMT high voltage and PMT offset: $(4.8\%, \SI{255.4}{\micro\meter}, 32, 0)$ in (a), $(0.3\%, \SI{255.4}{\micro\meter}, 1, 0)$ in (b), $(4.5\%, \SI{255.4}{\micro\meter}, 38, 0)$ in (c,e). Original image dimension is  \SI{151 x 151}{\micro\meter}, shown is zoom into data at \SI{100 x 100}{\micro\meter}.

\section*{Funding}
European Union Horizon 2020 research and innovation programme, Marie Skłodowska-Curie Actions (H2020-MSCA-IF-2014 655888), Beckman Institute Postdoctoral Fellowship from the Beckman Institute for Advanced Science and Technology, University of Illinois at Urbana-Champaign.

\section*{Acknowledgements}
We thank Prof. Brendan A. Harley and Prof. Paul J. Hergenrother at the University of Illinois at Urbana-Champaign (UIUC) for supplying Murine dental epithelial stem cells (mHAT9a) and Cellosaurus cell line (ES-2), respectively.

\section*{Disclosures}
MS and PSC are authors of US patent 9,213,313.

%%%%%%%%%% If using BibTeX:
%\bibliography{sample}
\bibliography{microscopy}

\begin{thebibliography}{10}
\newcommand{\enquote}[1]{``#1''}

\bibitem{RN7}
U.~Schnars and W.~Jüptner, \emph{Digital Holography: Digital Hologram
  Recording, Numerical Reconstruction, and Related Techniques} (Springer,
  2005).

\bibitem{RN66}
M.~K. Kim, \emph{Digital Holographic Microscopy: Principles, Techniques, and
  Applications} (Springer-Verlag, 2011).

\bibitem{RN217}
K.~Lee, K.~Kim, J.~Jung, J.~Heo, S.~Cho, S.~Lee, G.~Chang, Y.~Jo, H.~Park, and
  Y.~Park, \enquote{Quantitative phase imaging techniques for the study of cell
  pathophysiology: From principles to applications,}
  {\protect\JournalTitle{Sensors}} \textbf{13}, 4170 (2013).

\bibitem{RN9}
C.~Mann, L.~Yu, C.-M. Lo, and M.~Kim, \enquote{High-resolution quantitative
  phase-contrast microscopy by digital holography,} {\protect\JournalTitle{Opt.
  Express}} \textbf{13}, 8693--8698 (2005).

\bibitem{RN10}
B.~Rappaz, P.~Marquet, E.~Cuche, Y.~Emery, C.~Depeursinge, and P.~Magistretti,
  \enquote{Measurement of the integral refractive index and dynamic cell
  morphometry of living cells with digital holographic microscopy,}
  {\protect\JournalTitle{Opt. Express}} \textbf{13}, 9361--9373 (2005).

\bibitem{RN11}
T.~Zhang and I.~Yamaguchi, \enquote{Three-dimensional microscopy with
  phase-shifting digital holography,} {\protect\JournalTitle{Opt. Lett.}}
  \textbf{23}, 1221--1223 (1998).

\bibitem{RN12}
T.-C. Poon and T.~Kim, \enquote{Optical image recognition of three-dimensional
  objects,} {\protect\JournalTitle{Appl. Opt.}} \textbf{38}, 370--381 (1999).

\bibitem{RN13}
B.~W. Schilling, T.-C. Poon, G.~Indebetouw, B.~Storrie, K.~Shinoda, Y.~Suzuki,
  and M.~H. Wu, \enquote{Three-dimensional holographic fluorescence
  microscopy,} {\protect\JournalTitle{Opt. Lett.}} \textbf{22}, 1506--1508
  (1997).

\bibitem{RN14}
W.~S. Haddad, D.~Cullen, J.~C. Solem, J.~W. Longworth, A.~McPherson, K.~Boyer,
  and C.~K. Rhodes, \enquote{Fourier-transform holographic microscope,}
  {\protect\JournalTitle{Appl. Opt.}} \textbf{31}, 4973--4978 (1992).

\bibitem{RN15}
F.~Dubois, L.~Joannes, and J.-C. Legros, \enquote{Improved three-dimensional
  imaging with a digital holography microscope with a source of partial spatial
  coherence,} {\protect\JournalTitle{Appl. Opt.}} \textbf{38}, 7085--7094
  (1999).

\bibitem{RN1}
D.~Gabor, \enquote{A new microscopic principle,}
  {\protect\JournalTitle{Nature}} \textbf{161}, 777--778 (1948).

\bibitem{RN3}
E.~N. Leith and J.~Upatnieks, \enquote{Reconstructed wavefronts and
  communication theory,} {\protect\JournalTitle{J. Opt. Soc. Am.}} \textbf{52},
  1123--1128 (1962).

\bibitem{RN2}
P.~Hariharan, \emph{Basics of Holography} (Cambridge University Press, 2002).

\bibitem{RN219}
J.~Pawley, \emph{Handbook of Biological Confocal Microscopy} (Springer US,
  2006).

\bibitem{RN220}
D.~B. Murphy and M.~W. Davidson, \emph{Fundamentals of Light Microscopy and
  Electronic Imaging} (Wiley-Blackwell, 2013).

\bibitem{RN221}
H.~W. Robert, \enquote{Confocal optical microscopy,}
  {\protect\JournalTitle{Reports on Progress in Physics}} \textbf{59}, 427
  (1996).

\bibitem{RN225}
P.~Jacquemin, R.~McLeod, D.~Laurin, S.~Lai, and R.~A. Herring, \enquote{Design
  of a confocal holography microscope for three-dimensional temperature
  measurements of fluids in microgravity,} {\protect\JournalTitle{Microgravity
  Science and Technology}} \textbf{17}, 36--40 (2005).

\bibitem{RN37}
A.~S. Goy and D.~Psaltis, \enquote{Digital confocal microscope,}
  {\protect\JournalTitle{Opt. Express}} \textbf{20}, 22720--22727 (2012).

\bibitem{RN234}
A.~S. Goy, M.~Unser, and D.~Psaltis, \enquote{Multiple contrast metrics from
  the measurements of a digital confocal microscope,}
  {\protect\JournalTitle{Biomedical Optics Express}} \textbf{4}, 1091--1103
  (2013).

\bibitem{RN235}
C.~Liu, S.~Marchesini, and M.~K. Kim, \enquote{Quantitative phase-contrast
  confocal microscope,} {\protect\JournalTitle{Optics Express}} \textbf{22},
  17830--17839 (2014).

\bibitem{RN243}
N.~Lue, W.~Choi, K.~Badizadegan, R.~R. Dasari, M.~S. Feld, and G.~Popescu,
  \enquote{Confocal diffraction phase microscopy of live cells,}
  {\protect\JournalTitle{Optics Letters}} \textbf{33}, 2074--2076 (2008).

\bibitem{RN226}
R.~A. Herring, \enquote{Confocal scanning laser holography, and an associated
  microscope: A proposal,} {\protect\JournalTitle{Optik (Jena)}} \textbf{105},
  65--68 (1997).

\bibitem{RN223}
G.~E. Sommargren, \enquote{Optical heterodyne profilometry,}
  {\protect\JournalTitle{Applied Optics}} \textbf{20}, 610--618 (1981).

\bibitem{RN202}
C.~Rembe and A.~Dräbenstedt, \enquote{Laser-scanning confocal vibrometer
  microscope: Theory and experiments,} {\protect\JournalTitle{Review of
  Scientific Instruments}} \textbf{77}, 083702 (2006).

\bibitem{RN227}
H.~J. Matthews, D.~K. Hamilton, and C.~J.~R. Sheppard, \enquote{Surface
  profiling by phase-locked interferometry,} {\protect\JournalTitle{Applied
  Optics}} \textbf{25}, 2372--2374 (1986).

\bibitem{RN228}
B.~Deutsch, R.~Beams, and L.~Novotny, \enquote{Nanoparticle detection using
  dual-phase interferometry,} {\protect\JournalTitle{Applied Optics}}
  \textbf{49}, 4921--4925 (2010).

\bibitem{RN146}
J.~E. Graebner, \enquote{Optical scanning interferometer for dynamic imaging of
  high-frequency surface motion,} in \emph{Ultrasonics Symposium, 2000 IEEE,}
  vol.~1 (2000), pp. 733--736.

\bibitem{RN206}
J.~V. Knuuttila, P.~T. Tikka, and M.~M. Salomaa, \enquote{Scanning michelson
  interferometer for imaging surface acoustic wave fields,}
  {\protect\JournalTitle{Optics Letters}} \textbf{25}, 613--615 (2000).

\bibitem{RN224}
R.~J. Hocken, N.~Chakraborty, and C.~Brown, \enquote{Optical metrology of
  surfaces,} {\protect\JournalTitle{CIRP Annals}} \textbf{54}, 169--183 (2005).

\bibitem{RN176}
M.~Schnell, P.~S. Carney, and R.~Hillenbrand, \enquote{Synthetic optical
  holography for rapid nanoimaging,} {\protect\JournalTitle{Nat. Commun.}}
  \textbf{5}, 3499 (2014).

\bibitem{RN175}
M.~Schnell, M.~J. Perez-Roldan, P.~S. Carney, and R.~Hillenbrand,
  \enquote{Quantitative confocal phase imaging by synthetic optical
  holography,} {\protect\JournalTitle{Optics Express}} \textbf{22},
  15267--15276 (2014).

\bibitem{RN231}
A.~Di~Donato and M.~Farina, \enquote{Synthetic holography based on scanning
  microcavity,} {\protect\JournalTitle{AIP Advances}} \textbf{5}, 117125
  (2015).

\bibitem{RN232}
Y.~Zakharov, M.~Muravyeva, V.~Dudenkova, I.~Mukhina, E.~Vitkin, and
  L.~Perelman, \enquote{Holographic scanning microscopy – novel approach to
  digital holography and laser scanning microscopy,} in \emph{Imaging and
  Applied Optics 2014,}  (Optical Society of America, 2014), OSA Technical
  Digest (online), p. DW5B.1.

\bibitem{RN229}
C.~Liu, S.~Knitter, Z.~Cong, I.~Sencan, H.~Cao, and M.~A. Choma,
  \enquote{High-speed line-field confocal holographic microscope for
  quantitative phase imaging,} {\protect\JournalTitle{Optics Express}}
  \textbf{24}, 9251--9265 (2016).

\bibitem{RN244}
M.~Schnell, P.~S. Carney, and R.~Hillenbrand, \enquote{Transient vibration
  imaging with time-resolved synthetic holographic confocal microscopy,}
  {\protect\JournalTitle{Optics Express}} \textbf{26}, 26688--26699 (2018).

\bibitem{RN178}
B.~Deutsch, M.~Schnell, R.~Hillenbrand, and P.~S. Carney, \enquote{Synthetic
  optical holography with nonlinear-phase reference,}
  {\protect\JournalTitle{Optics Express}} \textbf{22}, 26621--26634 (2014).

\bibitem{RN265}
A.~Canales-Benavides, Y.~Zhuo, A.~M. Amitrano, M.~Kim, R.~I. Hernandez-Aranda,
  P.~S. Carney, and M.~Schnell, {\protect\JournalTitle{figshare}}  (2018).
  \url{https://osapublishing.figshare.com/s/4918dfa9f96f780e68ae}.

\bibitem{RN45}
N.~Ocelic, A.~Huber, and R.~Hillenbrand, \enquote{Pseudoheterodyne detection
  for background-free near-field spectroscopy,} {\protect\JournalTitle{Applied
  Physics Letters}} \textbf{89}, 101124 (2006).

\bibitem{RN239}
R.~Sekiya, M.~Maeda, H.~Yuan, E.~Asano, T.~Hyodo, H.~Hasegawa, S.~Ito,
  K.~Shibata, M.~Hamaguchi, F.~Kikkawa, H.~Kajiyama, and T.~Senga,
  \enquote{Plagl2 regulates actin cytoskeletal architecture and cell
  migration,} {\protect\JournalTitle{Carcinogenesis}} \textbf{35}, 1993--2001
  (2014).

\bibitem{RN238}
Y.~Zhuo, J.~S. Choi, T.~Marin, H.~Yu, B.~A. Harley, and B.~T. Cunningham,
  \enquote{Quantitative analysis of focal adhesion dynamics using photonic
  resonator outcoupler microscopy (prom),} {\protect\JournalTitle{Light:
  Science \& Applications}} \textbf{7}, 9 (2018).

\bibitem{RN242}
P.~Bon, S.~Lécart, E.~Fort, and S.~Lévêque-Fort, \enquote{Fast label-free
  cytoskeletal network imaging in living mammalian cells,}
  {\protect\JournalTitle{Biophysical Journal}} \textbf{106}, 1588--1595 (2014).

\bibitem{RN240}
C.~Edwards, B.~Bhaduri, T.~Nguyen, B.~G. Griffin, H.~Pham, T.~Kim, G.~Popescu,
  and L.~L. Goddard, \enquote{Effects of spatial coherence in diffraction phase
  microscopy,} {\protect\JournalTitle{Optics Express}} \textbf{22}, 5133--5146
  (2014).

\bibitem{RN241}
T.~H. Nguyen, C.~Edwards, L.~L. Goddard, and G.~Popescu, \enquote{Quantitative
  phase imaging with partially coherent illumination,}
  {\protect\JournalTitle{Optics Letters}} \textbf{39}, 5511--5514 (2014).

\bibitem{RN222}
T.~Sawatari, \enquote{Optical heterodyne scanning microscope,}
  {\protect\JournalTitle{Applied Optics}} \textbf{12}, 2768--2772 (1973).

\bibitem{RN262}
C.~J.~R. Sheppard, M.~Roy, and M.~D. Sharma, \enquote{Image formation in
  low-coherence and confocal interference microscopes,}
  {\protect\JournalTitle{Applied Optics}} \textbf{43}, 1493--1502 (2004).

\bibitem{RN260}
D.~Parshall and M.~K. Kim, \enquote{Digital holographic microscopy with
  dual-wavelength phase unwrapping,} {\protect\JournalTitle{Applied Optics}}
  \textbf{45}, 451--459 (2006).

\bibitem{RN261}
J.~Gass, A.~Dakoff, and M.~K. Kim, \enquote{Phase imaging without 2-pi
  ambiguity by multiwavelength digital holography,}
  {\protect\JournalTitle{Optics Letters}} \textbf{28}, 1141--1143 (2003).

\end{thebibliography}

%%%%%%%%%% If preparing manually:
% \newcommand{\enquote}[1]{``#1''}

% \bibitem{Zhang:14}
% Y.~Zhang, S.~Qiao, L.~Sun, Q.~W. Shi, W.~Huang, L.~Li, and Z.~Yang,
%   \enquote{Photoinduced active terahertz metamaterials with nanostructured
%   vanadium dioxide film deposited by sol-gel method,}
%   {\protect\JournalTitle{Optics Express}} \textbf{22}, 11070--11078 (2014).

% {Optical Society}, \enquote{{OSA Publishing},}
%   \url{http://www.osapublishing.org}.

% \bibitem{FORSTER2007}
% P.~Forster, V.~Ramaswamy, P.~Artaxo, T.~Bernsten, R.~Betts, D.~Fahey,
%   J.~Haywood, J.~Lean, D.~Lowe, G.~Myhre, J.~Nganga, R.~Prinn, G.~Raga,
%   M.~Schulz, and R.~V. Dorland, \enquote{Changes in atmospheric consituents and
%   in radiative forcing,} in \enquote{Climate Change 2007: The Physical Science
%   Basis. Contribution of Working Group 1 to the Fourth assesment report of
%   Intergovernmental Panel on Climate Change,}  S.~Solomon, D.~Qin, M.~Manning,
%   Z.~Chen, M.~Marquis, K.~B. Averyt, M.~Tignor, and H.~L. Miler, eds.
%   (Cambridge University Press, 2007).

% \end{thebibliography}

\end{document}